\documentclass[pra,twocolumn,showpacs,superscriptaddress,floatfix]{revtex4-2}
\usepackage{graphicx}
\usepackage{epstopdf} 
\usepackage{amsmath}
\usepackage{epsfig}
\usepackage{latexsym}
\usepackage{amsfonts}
\usepackage{graphics}
\usepackage{bm}
\usepackage{braket}
\usepackage{dsfont}
\usepackage{soul}
\usepackage{ulem}
\usepackage{bbm}

\usepackage{mathtools}

\usepackage{helvet}

\graphicspath{ {./Graphics/} }
\begin{document}

\date{\today}
\author{J. Mumford}
\affiliation{Department of Physics and Astronomy, McMaster University, 1280 Main St.\ W., Hamilton, ON, L8S 4M1, Canada}

\title{Meissner effect in Fock space}

\begin{abstract}
By periodically driving a single bosonic Josephson junction (BJJ) with an impurity, a synthetic gauge field is generated in the Fock space of the system.  At a critical synthetic gauge flux the ground state undergoes a quantum phase transition which is analogous to the Meissner-Abrikosov-vortex transition found in type-II superconductors with an applied magnetic field.  A second quantum phase transition involving attractive interactions between the bosons of the BJJ is shown to enhance the sensitivity of the system to the Meissner-Abrikosov-vortex transition.
\end{abstract}

\maketitle

\section{\label{Sec:Intro}Introduction}
Ultracold atomic gases have become the primary tool in simulating quantum phenomena \cite{jaksch05} due to their precise tunability and because they are among the simplest quantum many-body systems.    An exciting connection between ultracold atoms and condensed matter arises with the prospect of trapping the atoms in optical lattices because the spatial periodicity of the trapping potentials can be adjusted to produce a variety of crystal structures found in solids.  This has lead to the simulation of the motion of charged particles in materials, a prime example being the observation of the transition from a superfluid to a Mott insulator   \cite{greiner02}.  Simulating the motion of charged particles in magnetic fields has also been of interest due to the prospect of creating magnetic field strengths unattainable in conventional condensed matter experiments.  Synthetic gauge fields can be applied to the gas by periodically driving the external fields of the optical lattice and atoms \cite{dalibard11,goldman14a,goldman14b,eckardt15,burkov15,creffield16,aidelsburger11}.  The net result is a situation in which a particle can accumulate a non-zero phase in a closed path along the lattice mimicking the effect of a transverse magnetic flux.  This allows for the control of the topology of single-particle energy bands leading to quantum Hall physics and the realization of chiral edge states \cite{thouless82}.  Celebrated topological models have been simulated in this fashion like the Haldane \cite{haldane88,jotzu14,flaschner16} and Harper-Hofstadter \cite{harper55,aidelsburger13,miyake13} models.

The Meissner effect is another hallmark phenomenon in the interaction of magnetic fields and materials \cite{meissner33}.  In type-II superconductors, screening currents expel an external magnetic field by creating an opposing one.  When the magnetic field is larger than a critical value, $H > H_{c1}$, a quantum phase transition (QPT) occurs and the Meissner phase breaks down resulting in only partial screening of the field.  Vortices form at locations where the field penetrates the superconductor and is called the Abrikosov-vortex phase.  Theoretical interest in simulating this QPT has been primarily directed toward ladder systems as they are some of the simplest lattices which still allow for orbital magnetic field effects (a 1D chain lattice is insufficient).  Early investigations involved Josephson junction arrays to form the ladder \cite{kardar86,granato90,orignac01,denniston95} and more recently ultracold atoms in optical ladders have been proposed \cite{hugel14}.  Experimental verification came shortly after by loading a Bose-Einstein condensate (BEC) into an optical ladder and inducing a synthetic magnetic field through laser-assisted tunneling \cite{atala14}.  In the experiment, it was shown that chiral probability currents along each leg of the ladder play the role of the screening currents in the Meissner phase of a superconductor.  Past a critical flux of the synthetic field the leg currents decrease while the rung currents increase which was shown to be analogous to the Abrikosov-vortex phase.

In this work, we show that the Meissner-Abrikosov-vortex QPT can be realized with a BEC trapped in a double well potential, also known as a bosonic Josephson junction (BJJ).  The BEC is coupled to an impurity which can also tunnel between the two wells.  The array of Josephson junctions used in previous proposals can be reduced to a single BJJ by constructing the ladder lattice not in real space, but in Fock space called a Fock-state lattice (FSL) \cite{weiwang16,cai21,saugmann22,larson21,mumford22}.  The legs of the ladder are constructed from many-body states defined by the number of bosons of the BEC in each well and the rungs are constructed from the two similarly defined states of the impurity.  The synthetic magnetic field is generated by periodically driving the interaction strength between the BEC and the impurity as well as the tunneling between each well which effectively 'stirs' the FSL.  In experiments, both the boson-boson interactions of the BEC and the BEC-impurity interactions can be controlled via Feshbach resonance \cite{chin10} where the latter requires an ac-magnetic field to achieve periodic time dependence \cite{smith15}.  The double well potential is created by superimposing a 1D optical lattice onto a harmonic dipole trap, so the tunneling can be controlled by tuning the width and height of the optical barrier \cite{albiez05}.  Although we frame our analysis in terms of a BJJ, the two accessible states of the bosons and the impurity can be internal states as in the case of spinor condensates \cite{zibold10,gerving12}.  Furthermore, similar BEC-impurity interactions to the ones we use can be found in quantum dot systems such as a dipole-induced ion trapped between the two wells of a BJJ \cite{gerritsma12} and in solid state Josephson junctions \cite{pal18}.

The FSL is an example of the use of synthetic dimensions \cite{boada12,celi14,ozawa19b} which, in general, are degrees of freedom of a system which can imitate the motion of particles in real space.  In experiments, synthetic dimensions have been utilized with spin \cite{mancini15,stuhl15,anisimovas16}, momentum \cite{an17,meier16,xie19}, harmonic oscillator \cite{price17}, and rotational \cite{flob15} states of atoms and molecules.  When combined with synthetic gauge fields synthetic dimensions  have successfully simulated chiral edge states \cite{mancini15,stuhl15,kanungo22} and the measurement of Chern numbers of topological bands \cite{chalopin20}.  Usually a synthetic dimension is paired with at least one in real space, however, both dimensions in the FSL are synthetic which is a significant departure from the majority of experiments and proposals to date involving simulations with ultracold atoms.  

\begin{figure*}[t]
\centering
\includegraphics[scale=0.9]{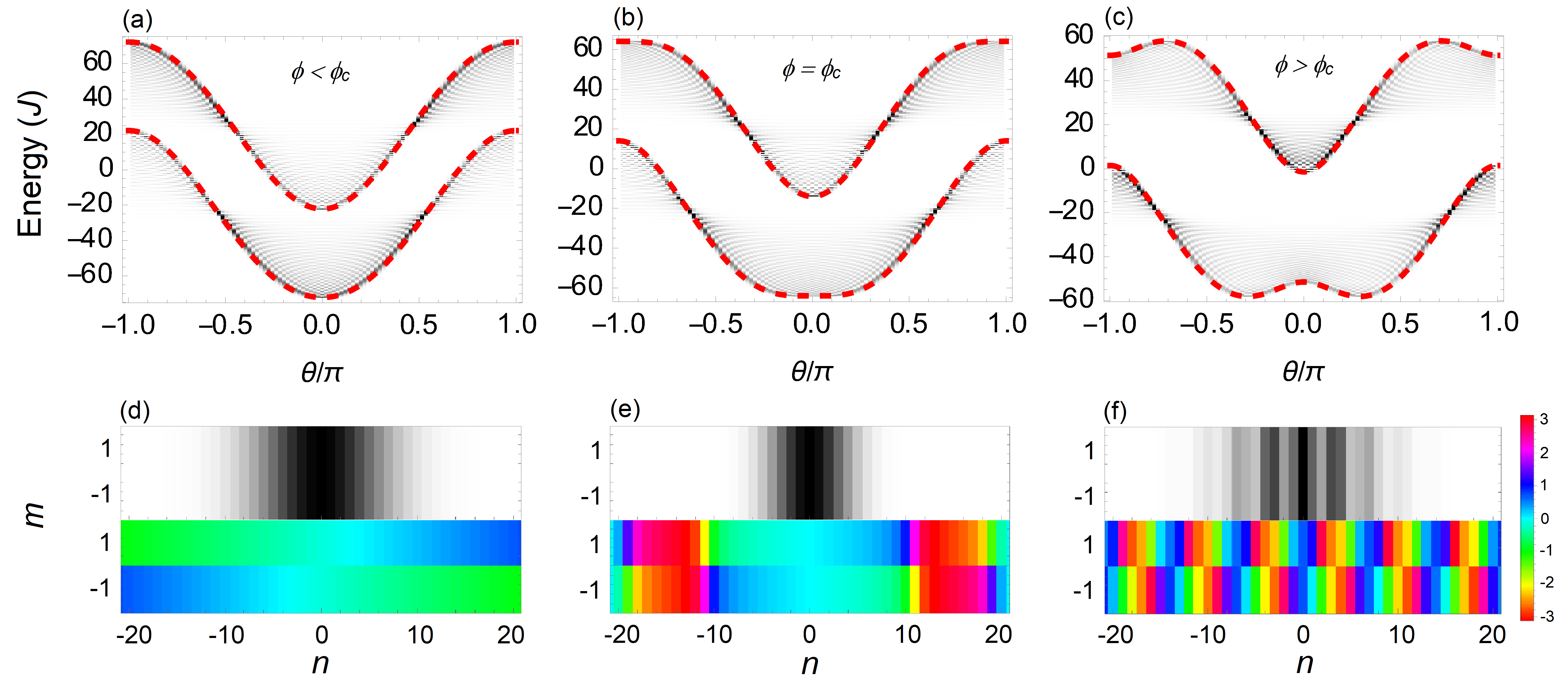}
\caption{Energy bands, ground state probability densities and phases in the FLS.  \textbf{(a)-(c)} Going from left to right the red dashed curves are the energy bands from Eq.\ \eqref{eq:enbands} for the magnetic fluxes $\phi_c/2$, $\phi_c$ and $3\phi_c/2$.  The background black and white density plots show the probability for a state with a given phase difference $\theta$ to have a given energy and are calculated using Eq.\ \eqref{eq:enphase}.  As the synthetic magnetic flux passes through the critical value, the lower band splits into a double well signifying a QPT where the ground state develops a non-zero doubly degenerate relative phase of the BEC.  \textbf{(d)-(f)} Each panel corresponds to the ground state probability distribution (top strips), $\vert \langle n \vert \epsilon_0 \rangle\vert^2$, and phase (bottom strips), $\mathrm{Arg}(\langle n \vert \epsilon_0 \rangle)$, of the lower band in the energy plots shown directly above.  In the Meissner phase the ground state is a Gaussian with a slowly varying phase along each leg of the ladder whereas in the vortex phase the wavefunction has holes in which vortices form which can be seen as a quickly varying phase in these regions.  The other parameter values are $N = 100$ and $\xi = 0.5$ and $\mu = 0$.}
\label{fig:en}
\end{figure*}

\section{\label{model}Model}

The system we will be investigating consists of a BEC coupled to an impurity in a double well potential.  Both of the BEC-impurity interactions and the tunneling between the two wells are periodically driven and are described by the Hamiltonian

\begin{equation}
\hat{H}(t) = U \hat{S}_z^2 + W \alpha(t) \hat{S}_z \hat{\sigma}_z - J \beta(t)\hat{S}_x - N K \gamma(t) \hat{\sigma}_x/2
\label{eq:ham}
\end{equation}
where $T$ is the period of the driving such that $\hat{H}(t) = \hat{H}(t+T)$.  The BEC and the impurity can occupy the ground states of each well, so every particle has access to two states which are labeled $L$ and $R$ for the left and right wells, respectively. In the Hamiltonian the creation and annihilation operators of the BEC are written in the Schwinger representation and are $\hat{S}_z = \frac{1}{2} (\hat{a}^\dagger_R\hat{a}_R - \hat{a}^\dagger_L \hat{a}_L )$, $\hat{S}_x = \frac{1}{2} (\hat{a}^\dagger_R\hat{a}_L + \hat{a}^\dagger_L \hat{a}_R )$ and $\hat{S}_y = \frac{\mathrm{i}}{2} (\hat{a}^\dagger_L\hat{a}_R - \hat{a}^\dagger_R \hat{a}_L )$ where the usual commutation relations are obeyed $[\hat{a}_i, \hat{a}^\dagger_j ] = \delta_{ij}$ where $i, j = L, R$.  The impurity operators are written as Pauli spin matrices.  The parameters are defined in the following way: $U$ is the boson-boson interaction energy of the BEC, $W$ is the BEC-impurity interaction energy, $J$ is the BEC tunneling energy, and $K$ is the impurity tunneling energy.  The factor of $N$ in front of the impurity tunneling term is the number of bosons in the BEC and is there to ensure this term does not become insignificant in the thermodynamic limit, $N \to \infty$.  The Greek letters represent the periodic driving and take the form

\begin{eqnarray}
\alpha(t) &=& - \sin(\omega t) \nonumber \\
\beta(t) &=& \sum_{p=0} \delta [t - (p+1/2)T] \nonumber \\
\gamma(t) &=& \sum_{p=0} \delta [t - pT]
\end{eqnarray}
which results in the BEC-impurity interactions being sinusoidally driven with frequency $\omega = 2\pi/T$ and the hopping terms being alternately pulsed in increments of $T/2$.  One cycle of the driving is achieved in four steps: (1) the impurity hopping is pulsed, (2) attractive BEC-impurity interactions are sinusoidally ramped up from, then back down to zero over an interval $T/2$, (3) the BEC tunneling is pulsed, (4)  repulsive BEC-impurity interactions are sinusoidally ramped up from, then back down to zero over an interval $T/2$.  The evolution after one cycle is given by the Floquet operator

\begin{eqnarray}
\hat{U}_F(t = T)= &&\mathrm{e}^{- \mathrm{i} \left [ \pi U \hat{S}_z^2 + 2 W \hat{S}_z \hat{\sigma}_z \right ] /\omega} \mathrm{e}^{ \mathrm{i} J \hat{S}_x \tau} \\ \nonumber 
&& \times \mathrm{e}^{- \mathrm{i} \left [ \pi U \hat{S}_z^2 - 2 W \hat{S}_z \hat{\sigma}_z \right ] /\omega} \mathrm{e}^{ \mathrm{i} NK \hat{\sigma}_x \tau/2 }
\end{eqnarray}
where $\tau \ll 1$ is the pulse interval.  When $T \ll 1$ the Floquet operator can be written in terms of an effective Hamiltonian $\hat{U}_F= \mathrm{e}^{-\mathrm{i} \hat{H}_\mathrm{eff} \tau}$ where

\begin{equation}
\hat{H}_\mathrm{eff}/J =2 \frac{\mu}{N} \hat{S}^2_z - \frac{1}{2} \left [ \hat{S}_+ \mathrm{e}^{-\mathrm{i} \phi \hat{\sigma}_z} + \hat{S}_- \mathrm{e}^{\mathrm{i} \phi \hat{\sigma}_z}  + N \xi \hat{\sigma}_x \right ] \, .
\label{eq:heff}
\end{equation}
The new parameters are defined as $\mu= \frac{ \pi U N}{J\omega \tau}$ and $\xi = K/J$.  The parameter $\phi = 2W/\omega$ appears as a Peierls phase seen in tight-binding models and so plays the role of the magnetic flux through the system.  The set of eigenvalues of $\hat{U}_F$, $\{ \lambda_i = \mathrm{e}^{\mathrm{i} \epsilon_i \tau} \}$, are defined in terms of the set of quasienergies $\{ \epsilon_i \}$.  The quasienergies are only unique in the range from $-\pi/\tau$ to $\pi/\tau$ because they are calculated from a unitary operator.  Therefore, the parameters in the model will be kept small enough to keep the quasienergies within this range. 

We assume BEC particle conservation, so the Hilbert space of the entire system has size $2\times \left ( N+1 \right )$.  A useful basis is the Fock basis made up of states with different particle number differences between the two wells.  With constant $N$ these states are labeled with a single number for both the BEC and the impurity $\vert n,m \rangle$ where $n = (N_R - N_L)/2$ is half of the BEC particle number difference between the two wells and $m = M_R - M_L$ is similarly the impurity number difference between the two wells.  We can extend the connections between this system and tight-binding models by considering the labels $n$ and $m$ as coordinates on a 2D FSL.  In this case, one dimension will be of size $N+1$ and the other will be of size $2$ forming a ladder lattice where each leg and each rung are formed by the BEC and impurity Fock states, respectively.  As an example, the Fock state $\vert 0, +1 \rangle$, which corresponds to an equal number of BEC particles in each well and the impurity in the right well, will be the coordinate of the center of the right leg of the ladder.

\section{\label{results}Results}

For now, we will assume the BEC is non-interacting and set $\mu = 0$.  We will also assume that we are in a regime with high coherence and low number difference fluctuations of the BEC, so that a mean-field approximation can be applied.  This amounts to replacing the BEC creation (annihilation) operators with complex numbers $\hat{a}_{R/L} = \sqrt{N_{R/L}} \mathrm{e}^{- \mathrm{i} \theta_{R/L}}$ where $\theta_{R/L}$ is the phase of the right/left wells.  The resulting mean-field Hamiltonian is 

\begin{eqnarray}
H_{\mathrm{MF}}/J =&&-\sqrt{N^2/4 - n^2}\cos\theta \cos\phi \nonumber \\
&& -  \sqrt{N^2/4 - n^2} \sin\theta \sin\phi \hat{\sigma}_z- \xi N\hat{\sigma}_x/2
\end{eqnarray}
where $\theta = \theta_R - \theta_L$ is the phase difference between the right and left wells (relative phase) and is the conjugate variable of $n$.  The final approximation we make is to assume that $n \ll N/2$, so that $\sqrt{N^2/4 - n^2} \approx N/2$.  The justification for this approximation comes from the fact that we are investigating a ground state QPT and the ground state of the BEC is a fairly localized state around $n = 0$.  The Hamiltonian becomes

\begin{equation}
H_{\mathrm{MF}}/J \approx -\frac{N}{2} [\cos\theta \cos\phi  +\sin\theta \sin\phi \hat{\sigma}_z + \xi \hat{\sigma}_x ] 
\label{eq:mf2}
\end{equation}
which resembles the Bloch Hamiltonian of a translationally invariant system with two sites per unit cell where $\theta$ plays the role of the quasimomentum.   The lattice spacing of the FSL is $a = 1$, so the range of the first Brillouin zone is $(-\pi, \pi)$ in increments of $\frac{2\pi}{N+1}$.  Such a system can be solved exactly giving the energies of the lower and upper bands

\begin{equation}
E_\pm(\theta)/J = -\frac{N}{2} \left [\cos\theta\cos\phi \mp \sqrt{\xi^2 + \sin^2\theta \sin^2\phi} \right ]
\label{eq:enbands}
\end{equation}
and eigenstates 

\begin{eqnarray}
\vert +, \theta \rangle &=& \cos(\alpha_\theta/2) \vert R, \theta \rangle -  \sin(\alpha_\theta/2) \vert L, \theta \rangle \nonumber \\
\vert -, \theta \rangle &=& \sin(\alpha_\theta/2) \vert R, \theta \rangle + \cos(\alpha_\theta/2) \vert L, \theta \rangle  
\label{eq:bands}
\end{eqnarray}
where 

\begin{equation}
\alpha_\theta = 2 \arctan \frac{1}{\xi}\left (\sin\theta \sin\phi + \sqrt{\xi^2 + \sin^2\theta \sin^2\phi}\right ) \, .
\label{eq:alpha}
\end{equation}

The system undergoes a QPT in the ground band which can be seen by finding the phases, $\theta_0$, that satisfy the condition $\partial E_-(\theta)/\partial \theta = 0$ giving

\begin{equation}
   \sin\theta_0 = 
\begin{cases}
  0 ,& \text{if } \phi \leq \phi_c\\
     \pm \sqrt{\sin^2\phi - \xi^2 \cot^2\phi},              & \text{if } \phi > \phi_c
\end{cases}
\label{eq:minphase}
\end{equation}
where the critical magnetic flux is 

\begin{equation}
\cos\phi_c = \frac{-\xi + \sqrt{\xi^2+4}}{2} \, .
\end{equation}
This QPT is from a zero phase difference below $\phi_c$ to a doubly degenerate phase difference above $\phi_c$.  The degeneracy comes from the spin-orbit coupling term in Eq.\ \eqref{eq:mf2}.  Its energy is minimized when the phase difference is negative (positive) and the the impurity is in the left (right) well, or using the FSL analogy, the left (right) legs of the ladder.  Putting it more succinctly, the degeneracy comes from $H_\mathrm{MF}$ being invariant under the flipping of both the relative phase of the BEC and the 'spin' of the impurity.  To get a visual sense of the transition and to check the validity of the approximations that have been made, in Fig.\ \ref{fig:en} (a)-(c) we plot the energies of each band derived in Eq.\ \eqref{eq:enbands} (dashed, red curves) as well as a density plot of the probability for a state with a given phase to occupy an eigenstate of the Floquet operator,

\begin{equation}
P_m(\theta, \epsilon_i) =\left \vert \sum_{n = -N/2}^{N/2} \mathrm{e}^{\mathrm{i} \theta n}\langle m, n \vert \epsilon_i \rangle \right  \vert^2 \, ,
\label{eq:enphase}
\end{equation}
for different values of $\phi$.  The agreement between the derived energy bands and the probability is excellent and shows that $\theta$ is a fairly good quantum number.  One can see that the effect of keeping the square root factors in the calculation of the probability is to cause it to fray away from the analytic result in certain areas.  The QPT can also be seen as the lower band goes from having a single minimum at $\theta = 0$ to having doubly degenerate non-zero minima.  

Figure  \ref{fig:en} (d)-(f) shows the probability distribution (top) and phase (bottom) of the ground state of $\hat{U}_F$, $\vert \epsilon_i \rangle$, in Fock space.  Note that the phase plotted is the argument of the wavefunction which is different from the phase difference between the two wells, $\theta$.  The first thing to notice is that the probability distribution is a Gaussian and is not spread over the entire FSL as one would expect for a translationally invariant system.  As previously mentioned, This confinement around $n \approx 0$ comes from the square root terms since their leading order $n$ dependence takes the form of a harmonic trap, $-\sqrt{N^2/4 - n^2} \approx -N/2 + n^2/N$.  For $\phi < \phi_c$ in panel (d) the phase of the ground state is slowly varying since $\theta_0 \approx 0$, however, for $\phi > \phi_c$ in panel (f) $\theta_0 \neq 0$ resulting in a sinusoidally varying phase in opposite directions along each leg of the ladder.  Vortices are located at the positions where the phases on each leg of the ladder differ the most (red/green) and create regions where the probability distribution almost vanishes.  The distribution does not completely go to zero, however, because positions in the system cannot be resolved beyond the lattice spacing, so vortex locations cannot be pinpointed exactly. 

\begin{figure}[t]
\centering
\includegraphics[width=\columnwidth]{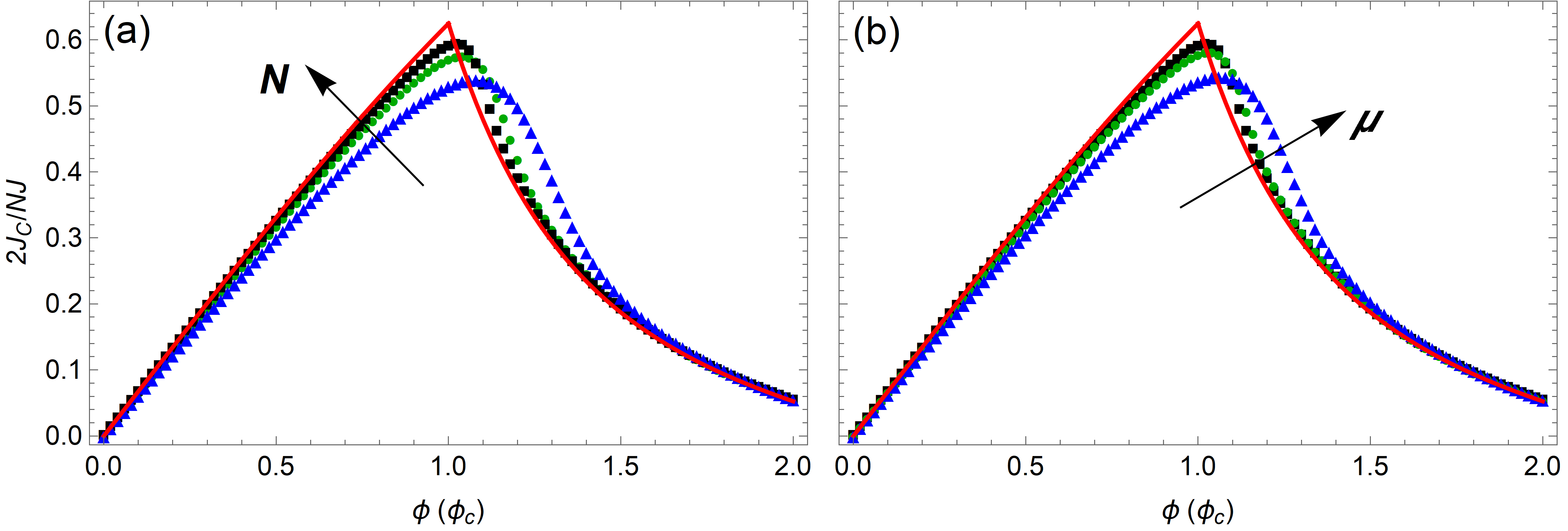}
\caption{Chiral current as a function of magnetic flux for different system sizes and boson-boson interaction energies. In both panels the red curves show the thermodynamic limit result from Eq.\ \eqref{eq:CC}.  \textbf{(a)} The chiral current is plotted for $N = 20$ (blue triangles), $N = 50$ (green circles) and $N = 100$ (black squares) and the other parameters values are $\xi = 0.5$, $\mu = 0$. As the system size increases the numerical results approach the analytic result because the larger the system size the more sensitive the system is to the presence of vortices. \textbf{(b)} The chiral current is plotted for $\mu = 5$ (blue triangles), $\mu = 0.5$ (green circles) and $\mu = 0$ (black squares) and other parameter values are $\xi = 0.5$ and $N = 100$.  As $\mu$ increases the peak moves away from the analytic result because the ground state wavefunction becomes more localized around $n = 0$ requiring larger fluxes to create a vortex with significant overlap of the wavefunction.}
\label{fig:CC}
\end{figure}

This QPT has the same qualitative features as the transition from the Meissner phase to the Abrikosov-vortex phase in type-II superconductors.  For this reason, we take a look at the probability current along the legs of the ladder.  The total current of the system is zero, however, the current difference between the two legs is non-zero and is called the chiral current \cite{greschner16}

\begin{equation}
J_C = J_R - J_L = \partial_\phi E_-(\theta)\vert_{\theta = \theta_0} \, .
\end{equation} 
On either side of the QPT the chiral current is expressed as 

\begin{equation}
   \frac{2J_C}{NJ} = 
\begin{cases}
  \sin(\phi) ,& \text{if } \phi \leq \phi_c\\
     \frac{\xi^2 \cos(\phi)}{\sin^2(\phi) \sqrt{\xi^2+\sin^2(\phi)}},              & \text{if } \phi > \phi_c \, .
\end{cases}
\label{eq:CC}
\end{equation}
Equation \eqref{eq:CC} is plotted as the red curves in Fig.\ \ref{fig:CC} (a) alongside the numerical result using the ground state of $\hat{U}_F$ for different system sizes.  The finite size scaling comes from competition between two lengthscales: the size of the probability distribution and the distance between vortices.  In panel (e) of Fig.\ \ref{fig:en} for $\phi = \phi_c$ vortices can be seen in the phase of the wavefunction, but they do not overlap much with the probability distribution, so they are insignificant.  The width of the probability distribution scales as $\sqrt{N}$, so for larger $N$ the easier it will be for a vortex to penetrate it.  In the thermodynamic limit, any presence of a vortex will be felt by the wavefunction and the numerical results will converge with the analytic result.

Repulsive boson-boson interactions have a similar affect on the probability distribution as the square root factors in that they act as a confining potential along the legs of the FSL.  This comes from the fact that there is an added energy cost for the BEC to occupy one well over the other and so there is an added energy cost for states $n \neq 0$ in the FSL.  The interaction energy, $\mu$, plays the role of the confining potential strength, so for stronger interactions larger fluxes are needed to produce a vortex that penetrates the wavefunction.  This leads to the peaks of the chiral currents appearing at larger flux values which can be seen  in panel (b) of Fig.\ \ref{fig:CC} where the chiral current is plotted for different values of $\mu$.

\begin{figure}[t]
\centering
\includegraphics[width=\columnwidth]{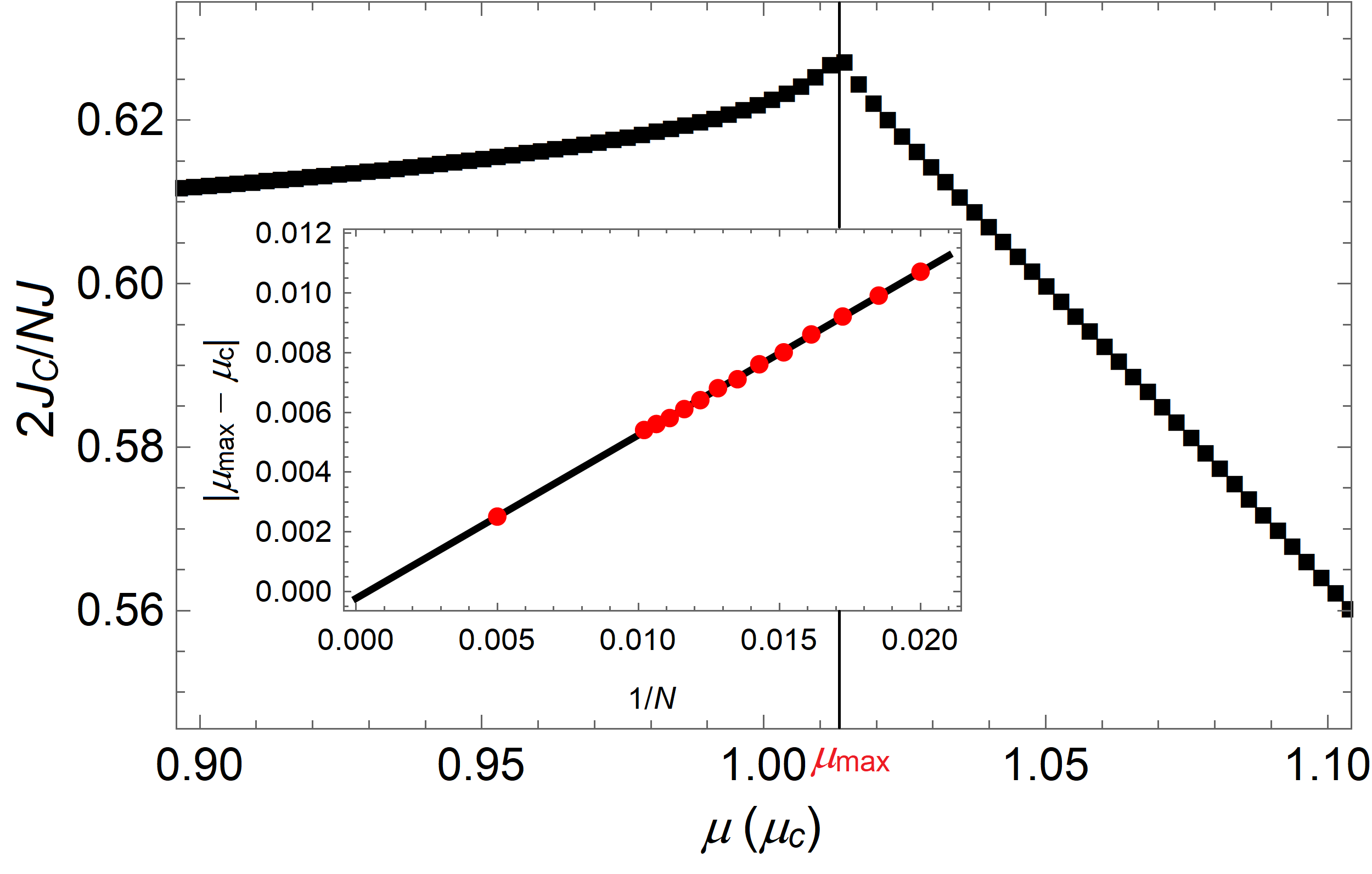}
\caption{The maximum chiral current as a function of $\mu$.  Each point of data represents the maximum chiral current of a $J_C$ vs. $\phi$ curve for a given value of $\mu$ like the ones seen in Fig.\ \ref{fig:CC} (b).  The value of $\mu$ where the peak occurs in this image is denoted as $\mu_\mathrm{max}$.  The inset shows $\vert \mu_\mathrm{max} - \mu_c \vert$ as a function of $1/N$ for different system sizes.  The black line is a linear fit and when extrapolated, shows that the difference goes to zero in the thermodynamic limit, $1/N \to 0$.}
\label{fig:JC}
\end{figure}

Conversely, attractive boson-boson interactions ($\mu < 0$) can enhance the Meissner-vortex QPT because they make it energetically favorable for the BEC to build-up in the wells rather that stay evenly dispersed as in the repulsive case.  This means the ground state wavefunction will be more spread out from the region around $n = 0$ and will be more susceptible to the presence of vortices.  In fact, in a normal BJJ without an impurity ($W = K = 0$), there is a critical interaction energy where the interactions overpower the confinement of the square root factors and another QPT occurs involving the breaking of a $\mathbb{Z}_2$ symmetry when the BEC spontaneously clumps into one well over the other \cite{buonsante12}.  At the critical point of this QPT, the correlation length in Fock space diverges and the width of the ground state wavefunction becomes comparable to $N$.  For a general magnetic flux, the interactions become dominant when $\mu < -\frac{1}{2} \cos\phi$.  Therefore, the Meissner-Abrikosov-vortex transition will be critically enhanced when the two QPTs coincide at an interaction energy of

\begin{equation}
\mu_c = -\frac{1}{2} \cos\phi_c = \frac{\xi - \sqrt{\xi^2+4}}{4} \, .
\label{eq:muc}
\end{equation}
To test this prediction, we calculate the maxima of $J_C$ vs. $\phi$ curves for different values of the interaction energy $\mu$.  The results for $N = 100$ are plotted in Fig.\ \ref{fig:JC} which shows there is a clear value of $\mu$, denoted as $\mu_\mathrm{max}$, where $J_C$ is a maximum.  The maximum occurs near $\mu_c$, but due to finite size effects, does not occur precisely at the critical point.  To show $\mu_\mathrm{max}$ converges to $\mu_c$ in the thermodynamic limit we plot their difference as a function of $1/N$ for different system sizes in the inset.  A linear fit is performed on the data and the resulting line is extrapolated to $1/N = 0$ giving a difference of $\vert \mu_\mathrm{max} - \mu_c \vert \approx 2.2 \times 10^{-4}$ at that point.  This provides numerical support to the result in Eq.\ \eqref{eq:muc} and one can conclude that the maximum possible chiral current occurs when the Meissner-Abrikosov-vortex QPT and the BEC clumping QPT coincide.

\section{Conclusion}

We have shown that FSLs offer a new direction to explore in the ever-growing field of quantum simulations of condensed matter systems.  By using a FSL, the usual techniques of simulating the Meissner effect with optical ladders and arrays of Josephson junctions is reduced to a single periodically driven BJJ coupled to an impurity.  The system exhibits the Meissner-Abrikosov-vortex QPT which we show can be enhanced by a second QPT involving the BEC spontaneously clumping into one well when the boson-boson interactions reach a critical attraction strength.  The Meissner-Abrikosov-vortex QPT is characterized by the chiral current, but a qualitative detection of the QPT can be achieved by measuring the relative phase of the BEC between the two wells since it transitions from zero in the Meissner phase to non-zero in the Abrikosov-vortex phase.  The relative phase can be measured from interference patterns in standard time-of-flight images after the trapping potential has been turned off \cite{andrews97,gati06}.

Although the system we have investigated is topologically trivial, additional BEC-impurity interactions can change that.  For instance, the inclusion of an impurity-assisted tunneling term like $\hat{S}_x \hat{\sigma}_x$ can result in a FSL version of the Su-Schrieffer-Heeger model.  Pairing this with the fact that a single impurity is enough to introduce a new synthetic dimension to the system opens up interesting possibilities for simulations of topological models in spaces higher than 2D.  Furthermore, since two particles are enough to form a 2D lattice and periodically driving their interactions generates the synthetic gauge field, this work opens the door for the simulation of many-body physics in the miniature.  The question then arises as to what is the smallest possible system one can use to simulate condensed matter phenomena of interest.

\appendix

\section{Numerical calculation of $J_C$}

To calculate the chiral current numerically we start with Eq.\ \eqref{eq:CC} and use the Hellmann-Feynman theorem to write 

\begin{equation}
J_C =  \langle \epsilon_0 \vert \partial_\phi \hat{H}_\mathrm{eff} \vert \epsilon_0 \rangle
\end{equation}
where $\vert \epsilon_0 \rangle$ is the ground state of $\hat{U}_F$.  The result of the derivative with respect to $\phi$ is

\begin{equation}
J_C = \langle \epsilon_0 \vert \hat{J}_x \vert \epsilon_0 \rangle \sin\phi - \langle \epsilon_0 \vert \hat{J}_y \hat{\sigma}_z \vert \epsilon_0 \rangle \cos\phi
\end{equation}
which is what is used in our calculations.

\section{Entanglement entropy ($\mu = 0$)}

\begin{figure}[t]
\centering
\includegraphics[width=\columnwidth]{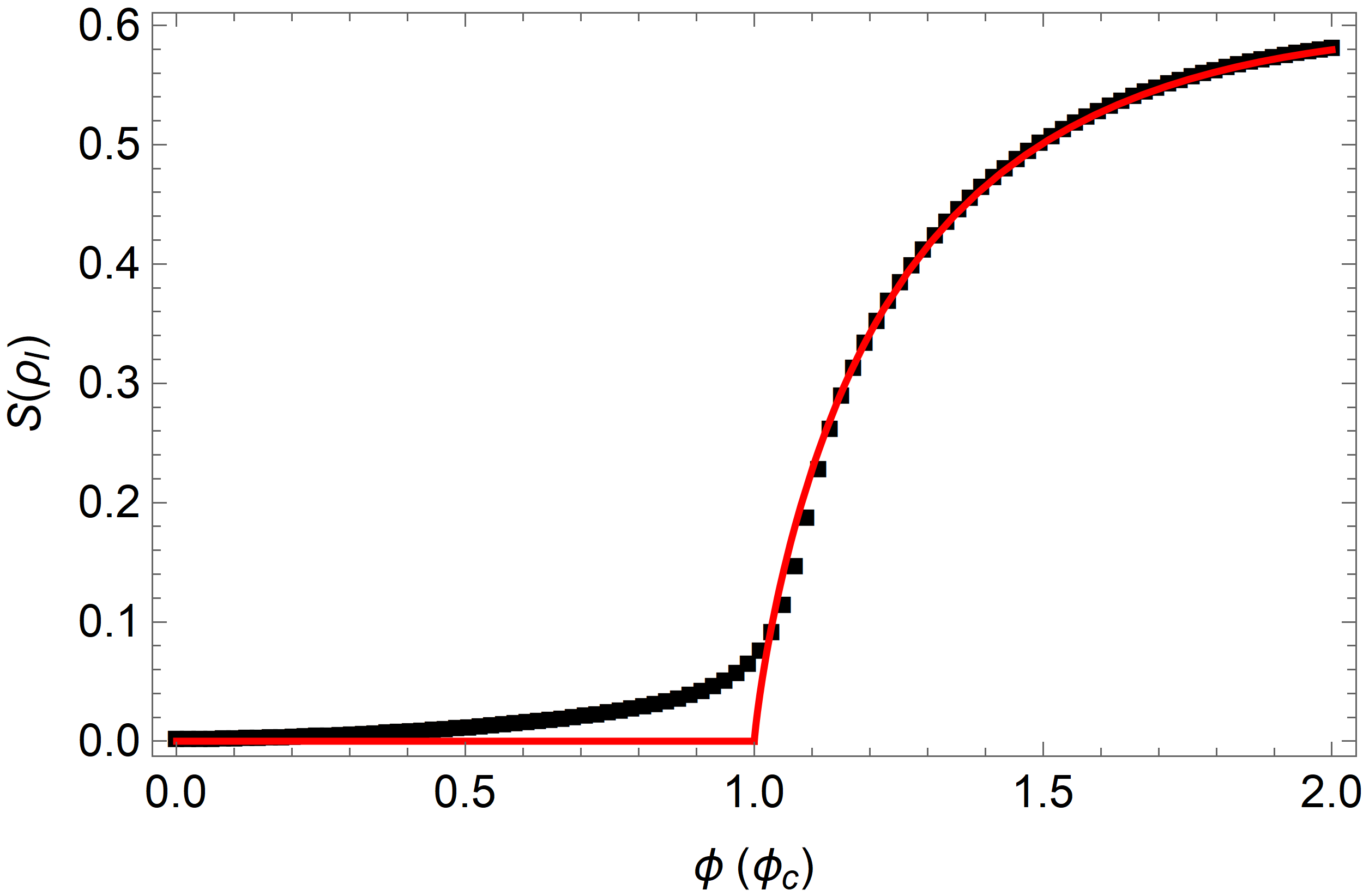}
\caption{The entanglement entropy between the BEC and the impurity.  The red curve is the analytic expression from Eq.\ \eqref{eq:analent} and the black squares are from a numerical calculation.  The parameters are $\xi = 0.5$, $\mu = 0$ and $N = 100$.}
\label{fig:EE}
\end{figure}

In the Meissner phase the screening current is entirely along the legs of the ladder (BEC degrees of freedom).  However, in the Abrikosov-vortex phase the current develops along the rungs of the ladder (impurity degrees of freedom) creating the vortices, so it is natural to ask whether or not the entanglement between the BEC and the impurity is sensitive to the QPT.  To check, we calculate the entanglement entropy of the ground state in Eq.\ \eqref{eq:bands} in both phases from the impurity reduced density matrix

\begin{equation}
\rho_\mathrm{I} = \vert -, \theta_0 \rangle \langle \theta_0, -\vert
\end{equation}
where $\theta_0$ is from Eq.\ \eqref{eq:minphase}.  The analytic expression for the entanglement entropy is

\begin{eqnarray}
S(\rho_\mathrm{I}) &=& - \mathrm{Tr}\left [\rho_\mathrm{I} \mathrm{ln}\rho_\mathrm{I} \right] \nonumber \\
&=& -f_+(\xi,\phi) \mathrm{ln}f_+(\xi,\phi) - f_-(\xi,\phi) \mathrm{ln}f_-(\xi,\phi)  \nonumber \\
\label{eq:analent}
\end{eqnarray}
where $f_\pm(\xi,\phi) = \frac{1}{2} \left (1 \pm \frac{\xi}{\sin\phi \sqrt{\xi^2 +\sin^2\phi}}\right )$.  The numerical calculation is carried out with the ground state of $\hat{U}_F$ giving the density matrix and impurity reduced density matrix

\begin{equation}
\rho = \vert \epsilon_0 \rangle \langle \epsilon_0 \vert , \hspace{10pt} \rho_\mathrm{I} = \mathrm{Tr}_\mathrm{B}\left [\rho \right]
\end{equation}
where $\mathrm{Tr}_\mathrm{B}$ is the trace over the BEC degrees of freedom.  The entanglement entropy is once again $S(\rho_\mathrm{I}) = - \mathrm{Tr}\left [\rho_\mathrm{I} \mathrm{ln}\rho_\mathrm{I} \right]$ which we compare to the analytic expression in Fig.\ \ref{fig:EE} and find excellent agreement between the two.  We see that, indeed, the entanglement entropy is sensitive to the QPT and follows the prediction of it increasing in the Abrikosov-vortex phase ($\phi > \phi_c$).  The disagreement between the analytic and numerical results around the critical point comes from finite size effects and is expected to vanish in the thermodynamic limit.

\end{document}